\documentclass{ws-procs9x6}

\def\bra{\langle}
\def\ket{\rangle}

\begin{document}

\title{COMPTON SCATTERING ON HE-3}

\author{D. CHOUDHURY$^*$ and D. R. PHILLIPS$^\dagger$}

\address{Department of Physics and Astronomy, Ohio University, Athens, Ohio 45701, USA\\
E-mail: $^*$choudhur@phy.ohiou.edu;
$^\dagger$phillips@phy.ohiou.edu}

\author{A. NOGGA}

\address{Institut f$\ddot{u}$r Kernphysik, Forschungszentrum J$\ddot{u}$lich, J$\ddot{u}$lich,
Germany\\
E-mail: a.nogga@fz-juelich.de}

\begin{abstract}
We present first calculations for Compton scattering on $^3$He. The
objective of the calculation is an extraction of the neutron
polarizabilities.
\end{abstract}

\bodymatter

\section{Introduction}\label{intro}

Our goal is to devise ways to extract the neutron polarizabilities.
Direct experiments on the neutron are not possible due to the lack
of free neutron targets and this encouraged physicists to look at
other avenues to extract information about the neutron
polarizabilities. Elastic Compton scattering on $^3$He is one such
avenue and here we report on the first calculations of this process.
Our results indicate that $\gamma \,^3$He scattering is indeed a
promising way to extract the neutron polarizabilities.

\section{The Calculation and Results}

The irreducible amplitudes for the elastic scattering of real
photons from the $NNN$ system are first ordered and calculated in
Heavy Baryon Chiral Perturbation Theory (HB$\chi$PT) to ${\mathcal
O}(Q^3)$-- they are the same as those computed in
Ref.~\refcite{Be99}. These amplitudes are then sandwiched between
the nuclear wavefunctions to finally obtain the scattering
amplitude--
\begin{equation}
{\mathcal M}=\bra \Psi_f|{\hat O}|\Psi_i \ket. \label{eq1}
\end{equation}
Using these amplitudes we calculate the differential cross-section
(dcs) or the double polarization observables, $\Delta_z$ or
$\Delta_x$~\cite{us}.

\begin{figure}
\begin{center}
\psfig{file=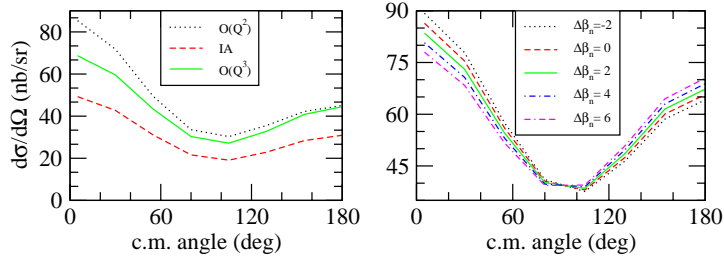,width=4in}
\end{center}
\caption{The differential cross-section vs. c.m. angle at 80~MeV.
Left panel shows $\frac{d\sigma}{d\Omega}$ calculated to different
orders. Right panel shows the sensitivity of
$\frac{d\sigma}{d\Omega}$ to $\Delta \beta_n$($\times
10^{-4}$fm$^3$).} \label{fig1}
\end{figure}

In Fig.~\ref{fig1} shows the differential cross-section in the
center of mass (c.m.) frame at 80~MeV. The left panel corresponds to
calculations at different orders-- ${\mathcal O}(Q^3)$, impulse
approximation (IA) which is actually ${\mathcal O}(Q^3)$ but does
not include two-body currents, and ${\mathcal O}(Q^3)$; and the
right panel to varying the value of $\beta_n$ around its ${\mathcal
O}(Q^3)$ predicted value.
\begin{figure}
\vspace{-0.2in}
\begin{center}
\psfig{file=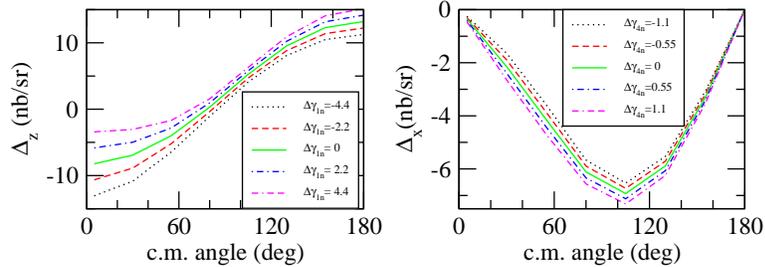,width=4in}
\end{center}
\caption{Figure showing the sensitivity of the double-polarization
observable, $\Delta_z$ to $\Delta \gamma_{1n}$($\times
10^{-4}$fm$^4$) and of $\Delta_x$ to $\Delta \gamma_{4n}$($\times
10^{-4}$fm$^4$) at 120~MeV in the c.m. frame.} \label{fig2}
\end{figure}
Fig.~\ref{fig2} shows the variation in $\Delta_z$ (at 120~MeV in the
c.m. frame) when $\Delta \gamma_{1n}$($\times 10^{-4}$fm$^4$) is
varied and that in $\Delta_x$ when $\Delta \gamma_{4n}$($\times
10^{-4}$fm$^4$) is varied.

Both Figs.~\ref{fig1} and \ref{fig2} suggest sizeable sensitivity to
the neutron polarizabilities in the $\gamma \,^3$He dcs and the
double polarization observables. For a more detailed description of
the calculation and results please refer to Ref.~\refcite{us}.

\end{document}